\documentclass[11pt,a4paper]{article}
\usepackage{jcappub}
\usepackage{float}
\usepackage{rotating}
\usepackage{graphicx}
\usepackage{epsfig}
\usepackage{amsmath}
\usepackage{amssymb}
\usepackage{mathrsfs}  
\usepackage{subfigure}
\usepackage{time}
\usepackage{orcidlink}
\usepackage{relsize, comment}
\usepackage{xcolor}
\usepackage[utf8]{inputenc}

\hypersetup{ linktoc=all,
    colorlinks=true, linkcolor={blue},  
       citecolor={red}, urlcolor={vividviolet}
}
\definecolor{rosy}{RGB}{230,235,252}
\definecolor{myframetitle}{RGB}{90,89,170}
\definecolor{myblocktitle}{RGB}{140,185,249}
\definecolor{mytitle}{RGB}{10,80,26}

\definecolor{darkgreen}{RGB}{27,130,45}
\definecolor{darkblue}{rgb}{0,0,0.3}
\definecolor{darkred}{rgb}{0.7,0,0}

\definecolor{light gray}{RGB}{220,220,220}
\definecolor{dark purple}{RGB}{108,0,217}
\definecolor{pink}{RGB}{190,20,100}
\definecolor{orang}{RGB}{193,63,0}
\definecolor{green}{RGB}{11,98,17}
\definecolor{darkpink}{RGB}{153,0,76}
\definecolor{bluegreen}{RGB}{0,102,102}
\definecolor{greenlagan}{RGB}{0,102,0}
\definecolor{redgreen}{RGB}{102,102,0}
\definecolor{Redgreen}{RGB}{153,76,0}
\definecolor{vividviolet}{rgb}{0.62, 0.0, 1.0}
\definecolor{amaranth}{rgb}{0.9, 0.17, 0.31}
\definecolor{palatinateblue}{rgb}{0.15, 0.23, 0.89}
\definecolor{brightpink}{rgb}{1.0, 0.0, 0.5}
\definecolor{cornflowerblue}{rgb}{0.39, 0.58, 0.93}
\definecolor{deepcarminepink}{rgb}{0.94, 0.19, 0.22}
\definecolor{radicalred}{rgb}{1.0, 0.21, 0.37}
\title{Lorentz Violation with Gravitational Waves: Constraints from NANOGrav and IPTA Data
}
\author[a]{Alireza Allahyari\orcidlink{0000-0002-4553-2436}}
\emailAdd{alireza.al@khu.ac.ir}
\author[a]{Mohammadreza Davari\orcidlink{0009-0001-8380-9328}}
\emailAdd{m.davari@khu.ac.ir}

\affiliation[a]{Department of Astronomy and High Energy Physics, Kharazmi University, 15719-14911, Tehran, Iran \looseness=-1}

\author[b]{and David F. Mota\orcidlink{0000-0003-3141-142X}}
\affiliation[b]{Institute of Theoretical Astrophysics, University of Oslo, P.O. Box 1029 Blindern, N-0315 Oslo, Norway}
\emailAdd{d.f.mota@astro.uio.no}

\abstract{
We explore a theoretical framework in which Lorentz symmetry is explicitly broken by incorporating derivative terms of the extrinsic curvature into the gravitational action. These modifications introduce a scale-dependent damping effect in the propagation of gravitational waves (GWs), governed by a characteristic energy scale denoted as \( M_{\text{LV}} \). We derive the modified spectral energy density of GWs within this model and confront it with recent observational data from the NANOGrav 15-year dataset and the second data release of the International Pulsar Timing Array (IPTA). Our analysis yields a lower bound on the Lorentz-violating energy scale, finding \( M_{\text{LV}} > 10^{-19} \) GeV at 68\% confidence level. This result significantly improves upon previous constraints derived from LIGO/VIRGO binary merger observations. Our findings demonstrate the potential of pulsar timing arrays to probe fundamental symmetries of spacetime and offer new insights into possible extensions of general relativity.
}

\begin{document}
\maketitle
\flushbottom
\section{Introduction}
%Lorentz symmetry is a fundamental symmetry in general relativity and standard model of particle physics. The violation of Lorentz symmetry can have profound implications for advancing our understanding of modern physics. It can lead to the discovery of physics beyond the standard model and indicate the existence of new theories or particles, or reveal unknown interactions or new fundamental structures in spacetime. Additionally, research on Lorentz symmetry violation can enhance quantum gravity theories and establish connections between general relativity and quantum mechanics.
Lorentz symmetry is a cornerstone of general relativity and the standard model of particle physics. At high-energy levels, it is widely believed that this symmetry will be broken. Its violation could reveal new physics beyond these frameworks, such as quantum gravity effects, variations in fundamental constants, or non-commutative geometry. Recent observations of gravitational waves (GWs) provide a unique opportunity to test Lorentz invariance at cosmological scales. In this work, we explore a model where Lorentz symmetry is broken by introducing extrinsic curvature derivatives into the gravitational action. Using data from NANOGrav and IPTA, we derive constraints on the energy scale of these modifications.

%There are various theoretical  modifications to GR. A crucial subset of these  theories has been the subject of significant attention for testing fundamental principles of GR, most importantly Lorentz invariance. At high-energy levels, it is widely believed that this invariance will be broken.

Precision tests have demonstrated that Lorentz symmetry and the associated CPT (Charge conjugation-Parity-Time reversal) symmetry are upheld with extraordinary accuracy \cite{Kostelecky:2008ts,Mattingly:2005re}. However, these tests have not covered all energy and length scales, nor have they exhausted the multitude of routes these symmetries could be violated \cite{Carenza:2025jwn}. Potential theories for Lorentz symmetry violation include spontaneous symmetry breaking in string theory \cite{Kostelecky:1988zi,Kostelecky:1991ak,Altschul:2005mu}, phenomena in loop quantum gravity \cite{Gambini:1998it,Alfaro:2001rb}, variations in fundamental constants over spacetime \cite{Kostelecky:2002ca,Ferrero:2009jb}, and non-commutative geometry \cite{Mocioiu:2000ip,Carroll:2001ws}, among others. Various modified gravity theories have been proposed to explore the nature of Lorentz violation including the Einstein-Æther theory \cite{Jacobson:2000xp, Eling:2004dk, Jacobson:2007veq, Li:2007vz, Zhang:2023kzs},
Horava-Lifshitz theories of quantum gravity \cite{Horava:2009uw, Takahashi:2009wc,  Wang:2012fi, Zhu:2013fja}, and
spatial covariant gravities \cite{Gao:2020qxy, Gao:2020yzr, Gao:2019twq}. Other studies consider various probes for Lorentz violation~\cite{ Kostelecky:2016kfm, Bailey:2006fd, Mewes:2019dhj, Shao:2020shv}.

%The gravitational waves observations has opened a new era in gravitational physics. When the Lorentz symmetry in gravity is broken, it could induce a different damping effect on gravitational wave propagation in GR.

Recently, strong indications of the possibility of stochastic gravitational waves background in nHz have been obtained by various pulsar timing arrays observations \cite{NANOGrav:2023gor,Reardon:2023gzh,EPTA:2023fyk,Xu:2023wog}. But there is still no precise description for the source of these waves \cite{NANOGrav:2023hvm}. Modifications to general relativity can significantly affect the spectrum of primordial gravitational waves. In this paper we find a lower bound on the main parameter of Lorentz violating $M_{LV}$ using NANOGrav 15 year \cite{NANOGrav:2023gor,NANOGrav:2023hvm} and IPTA second data release \cite{Perera:2019sca,Antoniadis:2022pcn}. For various recent works on Lorentz violation in connection with GWs see \cite{Zhang:2025kcw,Wang:2025fhw,Li:2024fxy,Zhang:2024rel,Hou:2024xbv,Amarilo:2023wpn,Ray:2023sbr,Zhu:2023rrx,Gong:2023ffb,Gong:2021jgg,Xu:2021dcw}.

Understanding Lorentz violation in the context of GWs is crucial for probing fundamental symmetries of nature and exploring potential modifications to general relativity. This work contributes to this effort by leveraging recent pulsar timing array (PTA) data to constrain Lorentz-violating effects.

In section (\ref{1}) we introduce the Lorentz violating damping effect for GWs. In section (\ref{2}) we derive the approximate transfer functions for tensor perturbations in Lorentz violating model and find the spectral energy density.  Finally, in section (\ref{3}) by using the NANOGrav and IPTA data, we constrain the model. Section~(\ref{4}) is summary and discussion.

\section{GWs in the Lorentz violating model  }\label{1}
In this section, we study a Lagrangian in which Lorentz invariance is broken and study the GWs equation in this theory.

The Lorentz violating damping effects can be introduced in the action by adding terms which contain extrinsic curvature derivatives like $\nabla_k K_{ij}$, $\nabla_k$ denotes the covariant derivative associated with the spatial metric $g_{ij}$.

More specifically,  we consider an example in which the mixed terms like $ \nabla_k K_{ij} \nabla^k K^{ij}$ are introduced. This construction  can arise in both the spatial covariant gravity \cite{Gao:2019liu} and Ho\v{r}ava-Lifshitz gravity \cite{Colombo:2014lta}.

By adding a  mixed term, which makes the theory power-counting renormalizable \cite{Zhu:2022uoq}, the  action can be written in ADM formalism as \cite{Zhang:2024rel}
\begin{eqnarray}
S &=& \frac{M_{\mathrm{Pl}}^{2}}{2} \int dtd^{3}x \sqrt{g} N (K_{ij} K^{ij} + R - K^{2}) \nonumber \\
&&+ \frac{M_{\mathrm{Pl}}^{2}}{2} \int dtd^{3}x \sqrt{g} N c_{1} (\nabla_{k}K_{ij}\nabla^{k}K^{ij} - R_{ij}R^{ij}), \nonumber \\ \label{action_ADM}
\end{eqnarray}
where $N$ is the lapse function. The first term represents the Einstein-Hilbert action  and the second term is the Lorentz violating modification. The coupling  $c_{1}$ is the coupling coefficient is a function of time $t$, and $M_{\rm Pl}$ is the reduced Planck mass.  Let us mention that 
$R_{ij}R^{ij}$  is used to  eliminate the effect of the  $\nabla_k K_{ij} \nabla^k K^{ij}$ in the dispersion relation for GWs. This way the theory predicts that the GWs propagate at the speed of light. 

We can expand the above action around a flat FRW background given by 
\begin{equation}
ds^2=-dt^2+a(t)^2\left\lbrace \delta_{ij}dx^{i}dx^j+h_{ij} dx^{i}dx^j  \right\rbrace \,,
\end{equation}
$h_{ij}$ denotes tensor perturbations. This way, action for GWs can be expanded up to the quadratic order as 
\begin{eqnarray}\label{quadratic}
S^{(2)} &=& \frac{M_{\rm Pl}^2}{8}\int dtd^{3}x a^{3} \left( \dot{h}_{ij}\dot{h}^{ij} +{h}_{ij}\frac{\triangle}{a^{2}}{h}^{ij} \right. \\
&&  \left.- c_1 \dot{h}_{ij}\frac{\triangle}{a^{2}}\dot{h}^{ij} +c_1 h_{ij}\frac{\triangle^2}{a^{2}} h^{ij}  \right).\\
\end{eqnarray}
where dot denotes derivative with respect to the cosmic time $t$ and $\triangle \equiv \delta^{ij}\partial_{i}\partial_{j}$.

We can derive the equations for GWs form action (\ref{quadratic}) with respect to $h_{ij}$. By using 
$a(\tau) d\tau=dt$, where $\tau$ is the conformal time, 
we find the equation of motion for $h_{ij}$ as 
\begin{eqnarray} 
&&\left(1-c_1\frac{\partial^{2}}{a^{2}}\right)h^{\prime \prime}_{ij} + \left[2\mathcal{H}-c_1'\frac{\partial^{2}}{a^{2}}\right]h^{\prime}_{ij}\nonumber\\
&&-\left(1-c_1\frac{\partial^{2}}{a^{2}}\right)\partial^{2}h_{ij} = 0.
\end{eqnarray}
Prime denotes derivative with respect to the conformal time and  $\mathcal{H}$ is the Hubble parameter in terms of conformal time. When $c_1=0$, the equations reduce to GR equations.
In the Fourier space, for each polarization $A$, the equation can be written as
%%%%%%%%%%%%%%%%%%%%%%%%%%%%%%%%%%%%%%%%%%%% 
\begin{eqnarray}
h^{\prime \prime}_{A} + (2 + {\bar \nu}) \mathcal{H} h^{\prime}_{A} + k^2 h_{A} = 0, \label{EoM_LV}
\end{eqnarray}
%%%%%%%%%%%%%%%%%%%%%%%%%%%%%%%%%%%%
where we have 
\begin{eqnarray}
{\cal H} \bar{\nu} = \left[\ln \left(1+c_1 \frac{k^2}{a^2}\right)\right]'\simeq \left(c_1 \frac{k^2}{a^2}\right)',
\end{eqnarray}
where the last equality holds when $c_1$ term is small.
Because $c_1$ has the dimensions $[energy]^{-2}$, we may parameterize  it in a more convenient form as
\begin{eqnarray}
c_1(\tau) = \frac{\alpha_{\bar \nu}(\tau)}{M_{\rm LV}^2}.
\end{eqnarray}
Now, $M_{\rm LV}$ has the dimension of energy and $\alpha_{\bar \nu}$ captures the time dependence of $c_1$. We set $\alpha_{\bar \nu}=1$ in the our analysis. The effect of deviations from standard relativity enhances in higher energy scales. In section~\ref{3}, we use the PTA data to find the energy scale where the modifications are important.
%%%%%%%%%%%%%%%%%%%%%%%%%%%%%%%%%%%%%%%%%%%%%%%%
%%%%%%%%%%%%%%%%%%%%%%%%%%%%%%%%%%%%%%%%%%%%%
\section{GWs stochastic background }\label{2}
In this section we derive the formula for transfer function of GWs in this theory and find the spectral energy density of gravitational waves.

The relevant quantity in PTA observations is the spectral  energy density.
The current spectral energy density of GWs  can be derived as  \cite{Caprini:2018mtu}
\begin{equation}
\label{spec}
\Omega_{GW}(k,\tau_0) = \frac{1}{12H_0^2}T'(k,\tau_0)]^2 P_t(k)\,,
\end{equation}
where $T(k,\tau)$ is the transfer function and $H_0$ is the Hubble constant. Transfer function describes the  evolution of GW modes after the modes re-enter the horizon. The quantity $P_t(k)$ is the primordial power spectrum of GWs at the end of the inflationary period written as in terms of a tensor amplitude $A$ and a tilt $n_t$ as 
\begin{equation}
P_t(k) = A(\frac{k}{k_{\star}})^{n_t}\,,
\end{equation}
where 	$k_{\star}$ is a pivot scale set as $k_{\star}=0.05\,{\rm Mpc}^{-1}$.

The transfer function may be written as \cite{Caprini:2018mtu,Kuroyanagi:2020sfw}
%%%%%%%%%%%%%%%%%%%%%%%%%%%%%%%%%
\begin{equation}
\label{transferMG}
T(k,\tau) = e^\mathcal{D}\,T(k,\tau)_{GR}\,,
\end{equation}
%%%%%%%%%%%%%%%%%%%%%%%%%%%%%%%%%%%
\begin{eqnarray}
\mathcal{D} &=& - \frac{1}{2} \int_{\tau_e}^{\tau} {\cal H}\bar \nu d\tau 
= - \frac{1}{2} \left[\alpha_{\bar \nu} \left(\frac{k}{a M_{\rm LV}}\right)^{2}\right]\Bigg|^{a_{\tau}}_{a_e}.
\end{eqnarray}

%%%%%%%%%%%%%%%%%%%%%%%%%%%%%%%%%%%%%%%%%%%%%%%%%%%%%%%%%%%%%%%%%%%%%
\begin{figure}[htbp]
	\centerline{\includegraphics[scale=1.0]{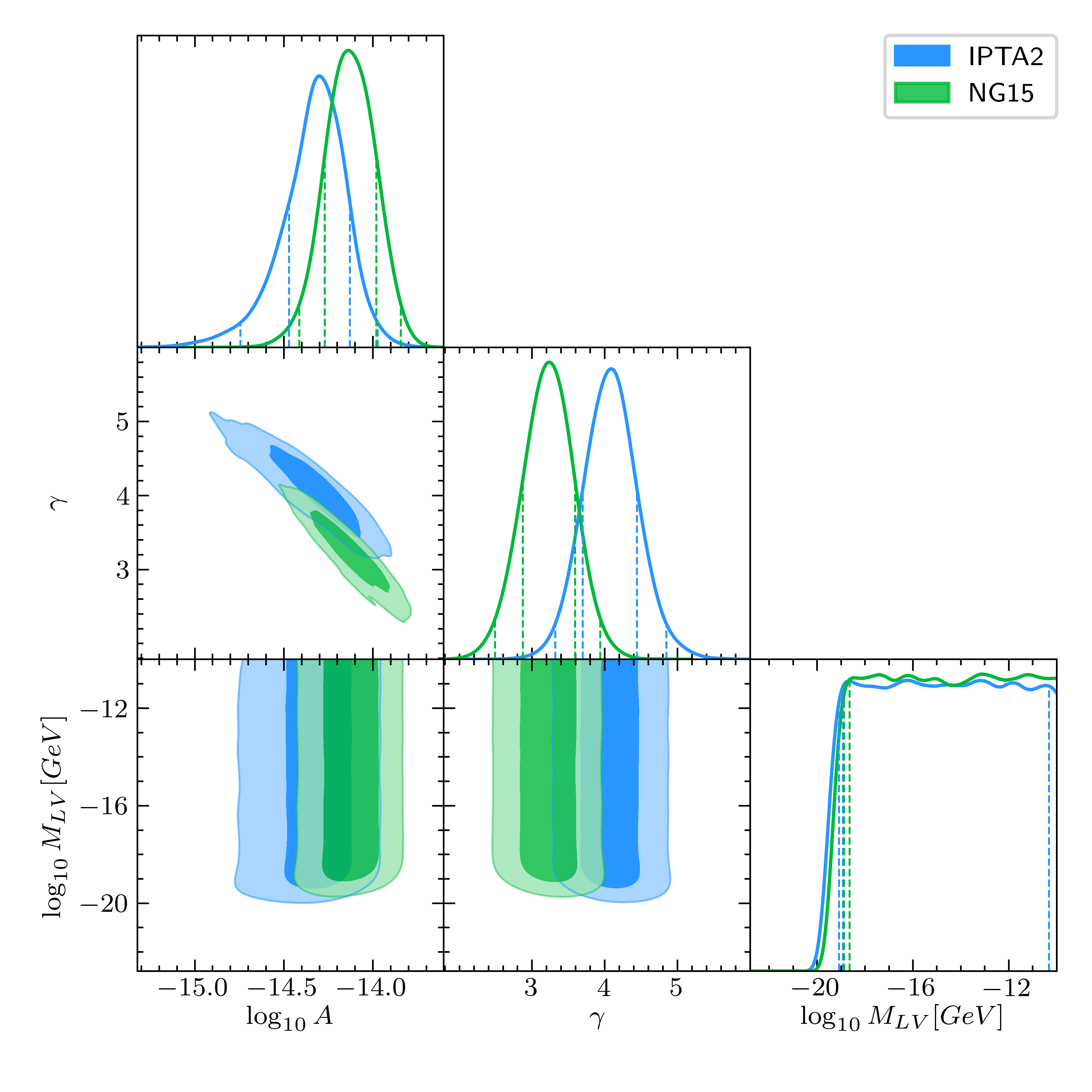}}
	\caption{Posterior distributions and marginal posteriors for the Lorentz-violating parameter $M_{LV}M_{LV}$, amplitude $AA$, and spectral index $\gamma\gamma$ using NANOGrav 15-year (NG15) and IPTA second data release (IPTA2). The contours represent 68\% and 95\% confidence levels.}
	\label{fig1}
\end{figure}
%%%%%%%%%%%%%%%%%%%%%%%%%%%%%%%%%%%%%%%%%%%%%%%%%%%%%%%%%%%%%%%%%%%%
where $a_e$ refers to the scale factor at horizon entry . To calculate this we set the parameters $H_0 = 67.36$ km/s/Mpc and $\Omega_r=9.2364 \times 10^{-5}$ from Planck 2018 \cite{Planck:2018vyg}. 
%So the relation of $a_e$ and the frequency of PTA %modes is
%\begin{equation}
%	a_e = \frac{H_0 \sqrt{\Omega_r}}{kc} \approx %\frac{3.34 \times 10^{-21}}{f}\,.
%\end{equation}
For GWs in PTA scales, we can use the approximation $(k \gg k_{eq})$ \cite{Vagnozzi:2023lwo}. 

In PTAs, it is convenient to express wavenumbers $k$ in terms of frequencies $f$ as \cite{Vagnozzi:2023lwo}  $f \simeq 1.54 \times 10^{-15} \left ( \frac{k}{{\rm Mpc}^{-1}} \right ) \,{\rm Hz}\,$.
We find that
$f_{\star} \simeq 7.7 \times 10^{-17}\,{\rm Hz}$.
Also in PTAs, the present  GWs spectral energy density is rather written in terms of the power spectrum of the GWs strain $h_c$ given by
\begin{eqnarray}
\Omega^{\rm PTA}_{\rm gw}(f) = \frac{2\pi^2}{3H_0^2}f^2h_c^2(f)\,.
\label{eq:omegagwhc}
\end{eqnarray}
$h_c(f)$ is supposed to take a power law form with respect to a reference frequency $f_{\rm yr}$ and can be expressed as
\begin{eqnarray}
h_c(f) = A \left ( \frac{f}{f_{\rm yr}} \right ) ^{\alpha}\,,
\label{eq:hc}
\end{eqnarray}
where $f_{\rm yr}=1\,{\rm yr}^{-1} \approx 3.17 \times 10^{-8}\,{\rm Hz}$.
Finally,  it is typical to write the current GWs spectral energy density by  introducing $\alpha = \frac{3-\gamma}{2}$. Then using the approximation $T'(k,\tau)_{GR}= k T(k,\tau)_{GR}$, We obtain the spectral energy density of GWs in the present time as

\begin{equation}
	\Omega_{GW}(f) = \frac{2\pi^2}{3H_0^2} A^2 \, e^{2\mathcal{D}}
	\left( \frac{\mathcal{D}' c}{2 \pi f} +1 \right)^2
	\left(\frac{f^{5-\gamma}}{yr^{\gamma-3}}\right)\,.
\end{equation}

\begin{equation}
\Omega_{GW}(f) = \frac{2\pi^2}{3H_0^2} A^2 \, \mathrm{exp}\left[(a_e^{-2} -1)\left((\frac{2\pi f}{c M_{LV}})^2\right)\right]
\left(\frac{2 \pi f H_0}{c M_{LV}^2} +1\right)^2
\left(\frac{f^{5-\gamma}}{yr^{\gamma-3}}\right)\,.
\end{equation}

%%%%%%%%%%%%%%%%%%%%%%%%%%%%
\section{Bounds on Lorentz violation by PTAs}\label{3}
In this section we use  NANOGrav 15 year data (NG15) and IPTA second data release (IPTA2) to find the constraints on $M_{LV}$. We use the python package \texttt{PTArcade} \cite{Mitridate:2023oar}. We consider uniform priors on the parameters as $-18<log_{10}A<-6,0<\gamma<6$ and $-25<\log_{10} M_{LV}<-10$. The results are illustrated in figure~\ref{fig1}.The best fits and the errors are also reported in table~\ref{tab1}.

We find that $M_{LV}> 10^{-19}\,{\rm GeV} $ at 68\% confidence. This result enhances the results obtained from binary mergers catalogs   $M_{LV}>10^{-21}\, {\rm GeV}$ by LIGO/VIRGO in \cite{Zhang:2024rel}. The value of $M_{LV}$ is affected by the value of $a_e$. Therefore, it is important to note that our result is obtained with $f=10^{-9}$ Hz. This is the lowest frequency in the PTA data. We checked that the higher the frequency, the higher the obtained bound will be, for instance, if we choose the reference frequency $f_{yr}$, the result will be about one order of magnitude better.
 \begin{table*}[htbp]
	\centering
	\begin{tabular}{|ccc|}
		\hline
		Parameter &  NG15 & IPTA2 \\
		\hline 		
		$\log_{10} M_{LV}[\rm GeV]$ & $>-19$ & $>-19$\\
		
		$log_{10}A$ & $-14.13\pm0.15$ & $-14.34^{+0.21}_{-0.14}$  \\
		
		$\gamma$ & $3.22\pm 0.37$ & $4.08\pm 0.39$\\
		\hline		
	\end{tabular}
	\caption{Constraints on the Lorentz-violating parameter $M_{LV}M_{LV}$, amplitude $AA$, and spectral index $\gamma\gamma$ at 68\% confidence level, derived from NANOGrav 15-year (NG15) and IPTA second data release (IPTA2).}
	\label{tab1}
\end{table*}

\begin{figure}[htbp]
	\centerline{\includegraphics[scale=1.0]{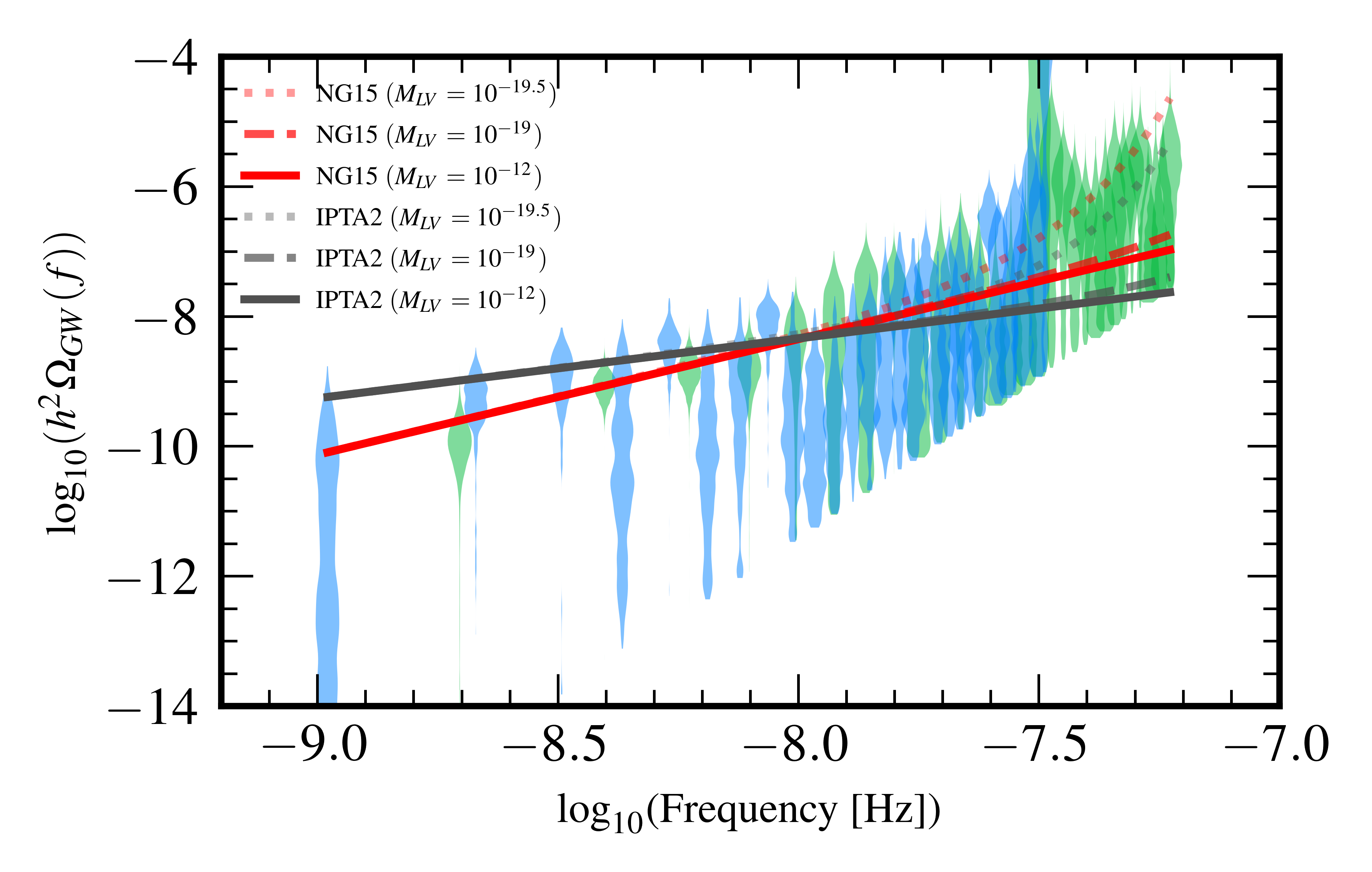}}
	\caption{The current spectral energy density of gravitational waves as a function of frequency. The violin plots show data from NANOGrav 15-year (NG15) and IPTA second data release (IPTA2), along with theoretical predictions for different values of $M_{LV}M_{LV}$.}
	\label{fig2}
\end{figure}

For concreteness, we show $h^2\Omega_{GW}$ as a function of frequency in logarithmic scales using the best fit of values of the parameters obtained from NG15 and IPTA2 in figure~\ref{fig2}. The violin plots are from NG 15-year and IPTA second data release.

%%%%%%%%%%%%%%%%%%%%%%%%%%%%%%%%%

%%%%%%%%%%%%%%%%%%%%%%%%%%%%%%%%%%%%%%%%%%%%%%%%%%%%%%%%%%%%%%%%%%%%%%%%%%%%%%
%%%%%%%%%%%%%%%%%%%%%%%%%%%%%%%%%%%%%%%%%%%%%%%%%%%%%%%%%%%%%%%%%%%%%%%%%%%%%%%
\section{Summary and discussion} \label{4}

The study investigates a theoretical model where Lorentz symmetry is broken by introducing terms with derivatives of the extrinsic curvature into the gravitational action. This modification alters the propagation of gravitational waves, introducing a scale-dependent damping effect. The energy scale of these Lorentz violating terms is parameterized by \( M_{LV} \). The analysis uses data from the \textbf{NANOGrav 15-year dataset (NG15)} and the \textbf{International Pulsar Timing Array second data release (IPTA2)} to constrain \( M_{LV} \). The study finds that \( M_{LV} > 10^{-19} \, \text{GeV} \) at  68\% confidence level. This result improves upon previous constraints from LIGO/VIRGO observations, which gave \( M_{LV} > 10^{-21} \, \text{GeV} \). The spectral energy density of gravitational waves is derived, incorporating the Lorentz violating damping effect. The modifications to the gravitational wave spectrum are significant at higher energy scales, as the damping effect becomes more pronounced. The posterior plots (Figure 1) show the constraints on \( M_{LV} \), the amplitude \( A \), and the spectral index \( \gamma \) for both NG15 and IPTA2 datasets. The best-fit values and uncertainties for these parameters are provided in Table 1: \( \log_{10} M_{LV} [\text{GeV}] > -19 \) for both NG15 and IPTA2, \( \log_{10} A \approx -14.13 \pm 0.15 \) (NG15) and \( -14.34^{+0.21}_{-0.14} \) (IPTA2), and \( \gamma \approx 3.22 \pm 0.37 \) (NG15) and \( 4.08 \pm 0.39 \) (IPTA2). The current spectral energy density of gravitational waves \( \Omega_{GW}(f) \) is plotted as a function of frequency (Figure 2). The violin plots in Figure 2 show the observed data from NG15 and IPTA2, along with theoretical predictions for different values of \( M_{LV} \). The constraints on \( M_{LV} \) depend on the frequency of the gravitational waves. Higher frequencies yield stronger constraints on \( M_{LV} \). For example, using a reference frequency \( f_{yr} \approx 3.17 \times 10^{-8} \, \text{Hz} \), the constraint on \( M_{LV} \) improves by about one order of magnitude. The study provides a framework for testing Lorentz violation using gravitational wave observations from pulsar timing arrays. The results suggest that Lorentz violating effects, if present, must occur at energy scales above \( 10^{-19} \, \text{GeV} \). 
%This work contributes to the broader effort to probe fundamental symmetries of nature and explore potential modifications to general relativity.

In summary, our study provides robust constraints on Lorentz-violating effects in the propagation of gravitational waves, using data from NANOGrav and IPTA. These results contribute to the broader effort to test fundamental symmetries of nature and explore potential modifications to general relativity. Future observations at higher frequencies and with improved sensitivity will further refine these constraints and deepen our understanding of gravity at cosmological scales.

\section{Acknowledgements}
DFM thanks the Research Council of Norway for their support and the resources provided by UNINETT Sigma2 -- the National Infrastructure for High-Performance Computing and Data Storage in Norway. 

\clearpage

\bibliographystyle{JHEP}
\bibliography{ref}

\providecommand{\href}[2]{#2}\begingroup\raggedright\begin{thebibliography}{10}

\bibitem{Kostelecky:2008ts}
V.~A. Kostelecky and N.~Russell, \emph{{Data Tables for Lorentz and CPT
  Violation}}, \href{https://doi.org/10.1103/RevModPhys.83.11}{\emph{Rev. Mod.
  Phys.} {\bfseries 83} (2011) 11--31},
  [\href{https://arxiv.org/abs/0801.0287}{{\ttfamily 0801.0287}}].

\bibitem{Mattingly:2005re}
D.~Mattingly, \emph{{Modern tests of Lorentz invariance}},
  \href{https://doi.org/10.12942/lrr-2005-5}{\emph{Living Rev. Rel.} {\bfseries
  8} (2005) 5}, [\href{https://arxiv.org/abs/gr-qc/0502097}{{\ttfamily
  gr-qc/0502097}}].

\bibitem{Carenza:2025jwn}
P.~Carenza, J.~Jaeckel, G.~Lucente, T.~K. Poddar, N.~Sherrill and
  M.~Spannowsky, \emph{{Limits on New Lorentz-violating Bosons}},
  \href{https://arxiv.org/abs/2502.05263}{{\ttfamily 2502.05263}}.

\bibitem{Kostelecky:1988zi}
V.~A. Kostelecky and S.~Samuel, \emph{{Spontaneous Breaking of Lorentz Symmetry
  in String Theory}},
  \href{https://doi.org/10.1103/PhysRevD.39.683}{\emph{Phys. Rev. D} {\bfseries
  39} (1989) 683}.

\bibitem{Kostelecky:1991ak}
V.~A. Kostelecky and R.~Potting, \emph{{CPT and strings}},
  \href{https://doi.org/10.1016/0550-3213(91)90071-5}{\emph{Nucl. Phys. B}
  {\bfseries 359} (1991) 545--570}.

\bibitem{Altschul:2005mu}
B.~Altschul and V.~A. Kostelecky, \emph{{Spontaneous Lorentz violation and
  nonpolynomial interactions}},
  \href{https://doi.org/10.1016/j.physletb.2005.09.018}{\emph{Phys. Lett. B}
  {\bfseries 628} (2005) 106--112},
  [\href{https://arxiv.org/abs/hep-th/0509068}{{\ttfamily hep-th/0509068}}].

\bibitem{Gambini:1998it}
R.~Gambini and J.~Pullin, \emph{{Nonstandard optics from quantum space-time}},
  \href{https://doi.org/10.1103/PhysRevD.59.124021}{\emph{Phys. Rev. D}
  {\bfseries 59} (1999) 124021},
  [\href{https://arxiv.org/abs/gr-qc/9809038}{{\ttfamily gr-qc/9809038}}].

\bibitem{Alfaro:2001rb}
J.~Alfaro, H.~A. Morales-Tecotl and L.~F. Urrutia, \emph{{Loop quantum gravity
  and light propagation}},
  \href{https://doi.org/10.1103/PhysRevD.65.103509}{\emph{Phys. Rev. D}
  {\bfseries 65} (2002) 103509},
  [\href{https://arxiv.org/abs/hep-th/0108061}{{\ttfamily hep-th/0108061}}].

\bibitem{Kostelecky:2002ca}
V.~A. Kostelecky, R.~Lehnert and M.~J. Perry, \emph{{Spacetime - varying
  couplings and Lorentz violation}},
  \href{https://doi.org/10.1103/PhysRevD.68.123511}{\emph{Phys. Rev. D}
  {\bfseries 68} (2003) 123511},
  [\href{https://arxiv.org/abs/astro-ph/0212003}{{\ttfamily
  astro-ph/0212003}}].

\bibitem{Ferrero:2009jb}
A.~Ferrero and B.~Altschul, \emph{{Radiatively Induced Lorentz and Gauge
  Symmetry Violation in Electrodynamics with Varying alpha}},
  \href{https://doi.org/10.1103/PhysRevD.80.125010}{\emph{Phys. Rev. D}
  {\bfseries 80} (2009) 125010},
  [\href{https://arxiv.org/abs/0910.5202}{{\ttfamily 0910.5202}}].

\bibitem{Mocioiu:2000ip}
I.~Mocioiu, M.~Pospelov and R.~Roiban, \emph{{Low-energy limits on the
  antisymmetric tensor field background on the brane and on the noncommutative
  scale}}, \href{https://doi.org/10.1016/S0370-2693(00)00928-X}{\emph{Phys.
  Lett. B} {\bfseries 489} (2000) 390--396},
  [\href{https://arxiv.org/abs/hep-ph/0005191}{{\ttfamily hep-ph/0005191}}].

\bibitem{Carroll:2001ws}
S.~M. Carroll, J.~A. Harvey, V.~A. Kostelecky, C.~D. Lane and T.~Okamoto,
  \emph{{Noncommutative field theory and Lorentz violation}},
  \href{https://doi.org/10.1103/PhysRevLett.87.141601}{\emph{Phys. Rev. Lett.}
  {\bfseries 87} (2001) 141601},
  [\href{https://arxiv.org/abs/hep-th/0105082}{{\ttfamily hep-th/0105082}}].

\bibitem{Jacobson:2000xp}
T.~Jacobson and D.~Mattingly, \emph{{Gravity with a dynamical preferred
  frame}}, \href{https://doi.org/10.1103/PhysRevD.64.024028}{\emph{Phys. Rev.
  D} {\bfseries 64} (2001) 024028},
  [\href{https://arxiv.org/abs/gr-qc/0007031}{{\ttfamily gr-qc/0007031}}].

\bibitem{Eling:2004dk}
C.~Eling, T.~Jacobson and D.~Mattingly, \emph{{Einstein-Aether theory}},  in
  \emph{{Deserfest: A Celebration of the Life and Works of Stanley Deser}},
  pp.~163--179, 10, 2004, \href{https://arxiv.org/abs/gr-qc/0410001}{{\ttfamily
  gr-qc/0410001}}.

\bibitem{Jacobson:2007veq}
T.~Jacobson, \emph{{Einstein-aether gravity: A Status report}},
  \href{https://doi.org/10.22323/1.043.0020}{\emph{PoS} {\bfseries QG-PH}
  (2007) 020}, [\href{https://arxiv.org/abs/0801.1547}{{\ttfamily 0801.1547}}].

\bibitem{Li:2007vz}
B.~Li, D.~Fonseca~Mota and J.~D. Barrow, \emph{{Detecting a Lorentz-Violating
  Field in Cosmology}},
  \href{https://doi.org/10.1103/PhysRevD.77.024032}{\emph{Phys. Rev. D}
  {\bfseries 77} (2008) 024032},
  [\href{https://arxiv.org/abs/0709.4581}{{\ttfamily 0709.4581}}].

\bibitem{Zhang:2023kzs}
C.~Zhang, A.~Wang and T.~Zhu, \emph{{Odd-parity perturbations of the
  wormhole-like geometries and quasi-normal modes in Einstein-\AE{}ther
  theory}}, \href{https://doi.org/10.1088/1475-7516/2023/05/059}{\emph{JCAP}
  {\bfseries 05} (2023) 059},
  [\href{https://arxiv.org/abs/2303.08399}{{\ttfamily 2303.08399}}].

\bibitem{Horava:2009uw}
P.~Horava, \emph{{Quantum Gravity at a Lifshitz Point}},
  \href{https://doi.org/10.1103/PhysRevD.79.084008}{\emph{Phys. Rev. D}
  {\bfseries 79} (2009) 084008},
  [\href{https://arxiv.org/abs/0901.3775}{{\ttfamily 0901.3775}}].

\bibitem{Takahashi:2009wc}
T.~Takahashi and J.~Soda, \emph{{Chiral Primordial Gravitational Waves from a
  Lifshitz Point}},
  \href{https://doi.org/10.1103/PhysRevLett.102.231301}{\emph{Phys. Rev. Lett.}
  {\bfseries 102} (2009) 231301},
  [\href{https://arxiv.org/abs/0904.0554}{{\ttfamily 0904.0554}}].

\bibitem{Wang:2012fi}
A.~Wang, Q.~Wu, W.~Zhao and T.~Zhu, \emph{{Polarizing primordial gravitational
  waves by parity violation}},
  \href{https://doi.org/10.1103/PhysRevD.87.103512}{\emph{Phys. Rev. D}
  {\bfseries 87} (2013) 103512},
  [\href{https://arxiv.org/abs/1208.5490}{{\ttfamily 1208.5490}}].

\bibitem{Zhu:2013fja}
T.~Zhu, W.~Zhao, Y.~Huang, A.~Wang and Q.~Wu, \emph{{Effects of parity
  violation on non-gaussianity of primordial gravitational waves in
  Ho\v{r}ava-Lifshitz gravity}},
  \href{https://doi.org/10.1103/PhysRevD.88.063508}{\emph{Phys. Rev. D}
  {\bfseries 88} (2013) 063508},
  [\href{https://arxiv.org/abs/1305.0600}{{\ttfamily 1305.0600}}].

\bibitem{Gao:2020qxy}
X.~Gao, \emph{{Higher derivative scalar-tensor theory from the spatially
  covariant gravity: a linear algebraic analysis}},
  \href{https://doi.org/10.1088/1475-7516/2020/11/004}{\emph{JCAP} {\bfseries
  11} (2020) 004}, [\href{https://arxiv.org/abs/2006.15633}{{\ttfamily
  2006.15633}}].

\bibitem{Gao:2020yzr}
X.~Gao and Y.-M. Hu, \emph{{Higher derivative scalar-tensor theory and
  spatially covariant gravity: the correspondence}},
  \href{https://doi.org/10.1103/PhysRevD.102.084006}{\emph{Phys. Rev. D}
  {\bfseries 102} (2020) 084006},
  [\href{https://arxiv.org/abs/2004.07752}{{\ttfamily 2004.07752}}].

\bibitem{Gao:2019twq}
X.~Gao and Z.-B. Yao, \emph{{Spatially covariant gravity theories with two
  tensorial degrees of freedom: the formalism}},
  \href{https://doi.org/10.1103/PhysRevD.101.064018}{\emph{Phys. Rev. D}
  {\bfseries 101} (2020) 064018},
  [\href{https://arxiv.org/abs/1910.13995}{{\ttfamily 1910.13995}}].

\bibitem{Kostelecky:2016kfm}
V.~A. Kosteleck\'y and M.~Mewes, \emph{{Testing local Lorentz invariance with
  gravitational waves}},
  \href{https://doi.org/10.1016/j.physletb.2016.04.040}{\emph{Phys. Lett. B}
  {\bfseries 757} (2016) 510--514},
  [\href{https://arxiv.org/abs/1602.04782}{{\ttfamily 1602.04782}}].

\bibitem{Bailey:2006fd}
Q.~G. Bailey and V.~A. Kostelecky, \emph{{Signals for Lorentz violation in
  post-Newtonian gravity}},
  \href{https://doi.org/10.1103/PhysRevD.74.045001}{\emph{Phys. Rev. D}
  {\bfseries 74} (2006) 045001},
  [\href{https://arxiv.org/abs/gr-qc/0603030}{{\ttfamily gr-qc/0603030}}].

\bibitem{Mewes:2019dhj}
M.~Mewes, \emph{{Signals for Lorentz violation in gravitational waves}},
  \href{https://doi.org/10.1103/PhysRevD.99.104062}{\emph{Phys. Rev. D}
  {\bfseries 99} (2019) 104062},
  [\href{https://arxiv.org/abs/1905.00409}{{\ttfamily 1905.00409}}].

\bibitem{Shao:2020shv}
L.~Shao, \emph{{Combined search for anisotropic birefringence in the
  gravitational-wave transient catalog GWTC-1}},
  \href{https://doi.org/10.1103/PhysRevD.101.104019}{\emph{Phys. Rev. D}
  {\bfseries 101} (2020) 104019},
  [\href{https://arxiv.org/abs/2002.01185}{{\ttfamily 2002.01185}}].

\bibitem{NANOGrav:2023gor}
{\scshape NANOGrav} collaboration, G.~Agazie et~al., \emph{{The NANOGrav 15 yr
  Data Set: Evidence for a Gravitational-wave Background}},
  \href{https://doi.org/10.3847/2041-8213/acdac6}{\emph{Astrophys. J. Lett.}
  {\bfseries 951} (2023) L8},
  [\href{https://arxiv.org/abs/2306.16213}{{\ttfamily 2306.16213}}].

\bibitem{Reardon:2023gzh}
D.~J. Reardon et~al., \emph{{Search for an Isotropic Gravitational-wave
  Background with the Parkes Pulsar Timing Array}},
  \href{https://doi.org/10.3847/2041-8213/acdd02}{\emph{Astrophys. J. Lett.}
  {\bfseries 951} (2023) L6},
  [\href{https://arxiv.org/abs/2306.16215}{{\ttfamily 2306.16215}}].

\bibitem{EPTA:2023fyk}
{\scshape EPTA, InPTA:} collaboration, J.~Antoniadis et~al., \emph{{The second
  data release from the European Pulsar Timing Array - III. Search for
  gravitational wave signals}},
  \href{https://doi.org/10.1051/0004-6361/202346844}{\emph{Astron. Astrophys.}
  {\bfseries 678} (2023) A50},
  [\href{https://arxiv.org/abs/2306.16214}{{\ttfamily 2306.16214}}].

\bibitem{Xu:2023wog}
H.~Xu et~al., \emph{{Searching for the Nano-Hertz Stochastic Gravitational Wave
  Background with the Chinese Pulsar Timing Array Data Release I}},
  \href{https://doi.org/10.1088/1674-4527/acdfa5}{\emph{Res. Astron.
  Astrophys.} {\bfseries 23} (2023) 075024},
  [\href{https://arxiv.org/abs/2306.16216}{{\ttfamily 2306.16216}}].

\bibitem{NANOGrav:2023hvm}
{\scshape NANOGrav} collaboration, A.~Afzal et~al., \emph{{The NANOGrav 15 yr
  Data Set: Search for Signals from New Physics}},
  \href{https://doi.org/10.3847/2041-8213/acdc91}{\emph{Astrophys. J. Lett.}
  {\bfseries 951} (2023) L11},
  [\href{https://arxiv.org/abs/2306.16219}{{\ttfamily 2306.16219}}].

\bibitem{Zhang:2025kcw}
B.-Y. Zhang, T.~Zhu, J.-M. Yan, J.-F. Zhang and X.~Zhang, \emph{{Constraining
  parity and Lorentz violations in gravity with future ground- and space-based
  gravitational wave detectors}},
  \href{https://arxiv.org/abs/2502.04776}{{\ttfamily 2502.04776}}.

\bibitem{Wang:2025fhw}
Q.~Wang, J.-M. Yan, T.~Zhu and W.~Zhao, \emph{{Modified gravitational wave
  propagations in linearized gravity with Lorentz and diffeomorphism violations
  and their gravitational wave constraints}},
  \href{https://arxiv.org/abs/2501.11956}{{\ttfamily 2501.11956}}.

\bibitem{Li:2024fxy}
T.-C. Li, T.~Zhu, W.~Zhao and A.~Wang, \emph{{Power spectra and circular
  polarization of primordial gravitational waves with parity and Lorentz
  violations}},
  \href{https://doi.org/10.1088/1475-7516/2024/07/005}{\emph{JCAP} {\bfseries
  07} (2024) 005}, [\href{https://arxiv.org/abs/2403.05841}{{\ttfamily
  2403.05841}}].

\bibitem{Zhang:2024rel}
B.-Y. Zhang, T.~Zhu, J.-F. Zhang and X.~Zhang, \emph{{Forecasts for
  constraining Lorentz-violating damping of gravitational waves from compact
  binary inspirals}},
  \href{https://doi.org/10.1103/PhysRevD.109.104022}{\emph{Phys. Rev. D}
  {\bfseries 109} (2024) 104022},
  [\href{https://arxiv.org/abs/2402.08240}{{\ttfamily 2402.08240}}].

\bibitem{Hou:2024xbv}
S.~Hou, X.-L. Fan, T.~Zhu and Z.-H. Zhu, \emph{{Nontensorial gravitational wave
  polarizations from the tensorial degrees of freedom: Linearized
  Lorentz-violating theory of gravity}},
  \href{https://doi.org/10.1103/PhysRevD.109.084011}{\emph{Phys. Rev. D}
  {\bfseries 109} (2024) 084011},
  [\href{https://arxiv.org/abs/2401.03474}{{\ttfamily 2401.03474}}].

\bibitem{Amarilo:2023wpn}
K.~M. Amarilo, M.~B.~F. Filho, A.~A.~A. Filho and J.~A. A.~S. Reis,
  \emph{{Gravitational waves effects in a Lorentz\textendash{}violating
  scenario}}, \href{https://doi.org/10.1016/j.physletb.2024.138785}{\emph{Phys.
  Lett. B} {\bfseries 855} (2024) 138785},
  [\href{https://arxiv.org/abs/2307.10937}{{\ttfamily 2307.10937}}].

\bibitem{Ray:2023sbr}
A.~Ray, P.~Fan, V.~F. He, M.~Bloom, S.~M. Yang, J.~D. Tasson et~al.,
  \emph{{Measuring gravitational wave speed and Lorentz violation with the
  first three gravitational-wave catalogs}},
  \href{https://doi.org/10.1103/PhysRevD.110.122001}{\emph{Phys. Rev. D}
  {\bfseries 110} (2024) 122001},
  [\href{https://arxiv.org/abs/2307.13099}{{\ttfamily 2307.13099}}].

\bibitem{Zhu:2023rrx}
T.~Zhu, W.~Zhao, J.-M. Yan, Y.-Z. Wang, C.~Gong and A.~Wang, \emph{{Constraints
  on parity and Lorentz violations in gravity from GWTC-3 through a
  parametrization of modified gravitational wave propagations}},
  \href{https://doi.org/10.1103/PhysRevD.110.064044}{\emph{Phys. Rev. D}
  {\bfseries 110} (2024) 064044},
  [\href{https://arxiv.org/abs/2304.09025}{{\ttfamily 2304.09025}}].

\bibitem{Gong:2023ffb}
C.~Gong, T.~Zhu, R.~Niu, Q.~Wu, J.-L. Cui, X.~Zhang et~al.,
  \emph{{Gravitational wave constraints on nonbirefringent dispersions of
  gravitational waves due to Lorentz violations with GWTC-3 events}},
  \href{https://doi.org/10.1103/PhysRevD.107.124015}{\emph{Phys. Rev. D}
  {\bfseries 107} (2023) 124015},
  [\href{https://arxiv.org/abs/2302.05077}{{\ttfamily 2302.05077}}].

\bibitem{Gong:2021jgg}
C.~Gong, T.~Zhu, R.~Niu, Q.~Wu, J.-L. Cui, X.~Zhang et~al.,
  \emph{{Gravitational wave constraints on Lorentz and parity violations in
  gravity: High-order spatial derivative cases}},
  \href{https://doi.org/10.1103/PhysRevD.105.044034}{\emph{Phys. Rev. D}
  {\bfseries 105} (2022) 044034},
  [\href{https://arxiv.org/abs/2112.06446}{{\ttfamily 2112.06446}}].

\bibitem{Xu:2021dcw}
R.~Xu, Y.~Gao and L.~Shao, \emph{{Signatures of Lorentz Violation in Continuous
  Gravitational-Wave Spectra of Ellipsoidal Neutron Stars}},
  \href{https://doi.org/10.3390/galaxies9010012}{\emph{Galaxies} {\bfseries 9}
  (2021) 12}, [\href{https://arxiv.org/abs/2101.09431}{{\ttfamily
  2101.09431}}].

\bibitem{Perera:2019sca}
B.~B.~P. Perera et~al., \emph{{The International Pulsar Timing Array: Second
  data release}}, \href{https://doi.org/10.1093/mnras/stz2857}{\emph{Mon. Not.
  Roy. Astron. Soc.} {\bfseries 490} (2019) 4666--4687},
  [\href{https://arxiv.org/abs/1909.04534}{{\ttfamily 1909.04534}}].

\bibitem{Antoniadis:2022pcn}
J.~Antoniadis et~al., \emph{{The International Pulsar Timing Array second data
  release: Search for an isotropic gravitational wave background}},
  \href{https://doi.org/10.1093/mnras/stab3418}{\emph{Mon. Not. Roy. Astron.
  Soc.} {\bfseries 510} (2022) 4873--4887},
  [\href{https://arxiv.org/abs/2201.03980}{{\ttfamily 2201.03980}}].

\bibitem{Gao:2019liu}
X.~Gao and X.-Y. Hong, \emph{{Propagation of gravitational waves in a
  cosmological background}},
  \href{https://doi.org/10.1103/PhysRevD.101.064057}{\emph{Phys. Rev. D}
  {\bfseries 101} (2020) 064057},
  [\href{https://arxiv.org/abs/1906.07131}{{\ttfamily 1906.07131}}].

\bibitem{Colombo:2014lta}
M.~Colombo, A.~E. Gumrukcuoglu and T.~P. Sotiriou, \emph{{Ho\v{r}ava gravity
  with mixed derivative terms}},
  \href{https://doi.org/10.1103/PhysRevD.91.044021}{\emph{Phys. Rev. D}
  {\bfseries 91} (2015) 044021},
  [\href{https://arxiv.org/abs/1410.6360}{{\ttfamily 1410.6360}}].

\bibitem{Zhu:2022uoq}
T.~Zhu, W.~Zhao and A.~Wang, \emph{{Gravitational wave constraints on spatial
  covariant gravities}},
  \href{https://doi.org/10.1103/PhysRevD.107.044051}{\emph{Phys. Rev. D}
  {\bfseries 107} (2023) 044051},
  [\href{https://arxiv.org/abs/2211.04711}{{\ttfamily 2211.04711}}].

\bibitem{Caprini:2018mtu}
C.~Caprini and D.~G. Figueroa, \emph{{Cosmological Backgrounds of Gravitational
  Waves}}, \href{https://doi.org/10.1088/1361-6382/aac608}{\emph{Class. Quant.
  Grav.} {\bfseries 35} (2018) 163001},
  [\href{https://arxiv.org/abs/1801.04268}{{\ttfamily 1801.04268}}].

\bibitem{Kuroyanagi:2020sfw}
S.~Kuroyanagi, T.~Takahashi and S.~Yokoyama, \emph{{Blue-tilted inflationary
  tensor spectrum and reheating in the light of NANOGrav results}},
  \href{https://doi.org/10.1088/1475-7516/2021/01/071}{\emph{JCAP} {\bfseries
  01} (2021) 071}, [\href{https://arxiv.org/abs/2011.03323}{{\ttfamily
  2011.03323}}].

\bibitem{Planck:2018vyg}
{\scshape Planck} collaboration, N.~Aghanim et~al., \emph{{Planck 2018 results.
  VI. Cosmological parameters}},
  \href{https://doi.org/10.1051/0004-6361/201833910}{\emph{Astron. Astrophys.}
  {\bfseries 641} (2020) A6},
  [\href{https://arxiv.org/abs/1807.06209}{{\ttfamily 1807.06209}}].

\bibitem{Vagnozzi:2023lwo}
S.~Vagnozzi, \emph{{Inflationary interpretation of the stochastic gravitational
  wave background signal detected by pulsar timing array experiments}},
  \href{https://doi.org/10.1016/j.jheap.2023.07.001}{\emph{JHEAp} {\bfseries
  39} (2023) 81--98}, [\href{https://arxiv.org/abs/2306.16912}{{\ttfamily
  2306.16912}}].

\bibitem{Mitridate:2023oar}
A.~Mitridate, D.~Wright, R.~von Eckardstein, T.~Schr\"oder, J.~Nay, K.~Olum
  et~al., \emph{{PTArcade}},
  \href{https://arxiv.org/abs/2306.16377}{{\ttfamily 2306.16377}}.

\end{thebibliography}\endgroup

\end{document}